\long\def\equalign
#1\end#2{\def\\{\cr\noalign{\smallskip}}
    \refstepcounter{equation}$$ \vcenter{\tabskip=5pt
    \halign{\hfil$\displaystyle{##}$&&$\displaystyle{{}##}$\hfil\cr
    #1 \cr}}\end{#2}}

\def\d3p{d^3p}
\def\tp3{(2\pi)^3}

\def\fq{f_q(p,t)}
\def\fm{f_{\pi}(p,t)}
\def\fe{f_{em}(p,t)}

\def\eq{f_{q, \,th}(p,t)}
\def\em{f_{\pi, \,th}(p,t)}

\def\tth{\tau_{th}(p,t)}
\def\thd{\tau_{had}(p,t)}
\def\tdc{\tau_{dec}(p,t)}
\def\tem{\tau_{em}(p)}

\def\vq{{\bf q} \,}
\def\vp{{\bf p} \,}
\def\vr{{\bf r} \,}
\def\vrp{{\bf r}^{\prime}\,}

\documentstyle[12pt]{article}

\begin{document}
\fnsymbol{footnote}
\hspace{11cm} HD-TVP-94-18 \\[5mm]

\begin{center}
\begin{large}                            
\bf{CRITICAL SCATTERING AT THE CHIRAL PHASE TRANSITION
    AND LOW-$p_T$ ENHANCEMENT OF MESONS IN
    ULTRA-RELATIVISTIC HEAVY-ION COLLISIONS} ${}^1$ \\[12mm]
\end{large}

Ji\v r\'{\i} Dolej\v s\'{\i} $^{2}$, Wojciech Florkowski $^{3}$,
and  J\"org H\"ufner \\[7mm]

Institut f\"ur Theoretische Physik, Universit\"at Heidelberg\\
D-69-120 Heidelberg, Philosophenweg 19, Germany\\[7mm]

\end{center}

{\bf Abstract:} The enhancement of
pions and kaons observed at small transverse momenta in
ultra-relativistic heavy-ion collisions may at least
partly reflect critical scattering expected to occur
in the neighborhood of a second order phase transition.
Kinetic equations in the relaxation time approximation are
proposed for the time evolution of the quark distribution function
into that of the pions. Relaxation times for thermalization
and hadronization processes are functions of momenta
and approach zero in the limit $p \rightarrow 0$, a consequence
of criticality at the phase transition. Data can be
reproduced for suitably chosen parameters.

\bigskip
\begin{center}
HEIDELBERG, OCTOBER 1994
\end{center}

\bigskip
\begin{itemize}
\item[${}^1$] Research supported in part by the Federal Minister
for Research and Technology (BMFT) grant number 06 HD 742 (0),
and the Deutsche Forschungsgemeinschaft, grant number Hu 233/4-3.
\item[${}^2$] Permanent address: Dept. of Nuclear Physics,
Charles University, V Hole\v sovi\v ck\' ach~2,
CZ-180~00~~Prague~8, Czech Republic.
\item[${}^3$] Permanent address: H. Niewodnicza\'nski Institute
of Nuclear Physics, ul. Radziko{-}wskiego 152, PL-31-342 Krak\'ow,
Poland.
\end{itemize}

\newpage
\renewcommand{\headheight}{-10mm}   
{\noindent \large \bf Introduction}
\medskip

The {\it low-$p_T$ enhancement} observed in ultra-relativistic
heavy-ion collisions is one of the unexpected results in
this domain of physics (c.f. the review \cite{SS}): At  small
values of the transverse mass, $m_T= \sqrt{m^2+p_T^2}$, the
observed distributions of pions and kaons, $dN/dm_T$, considerably
exceed the thermal distribution $\exp(-m_T/T)$
\cite{QM}. Various explanations have been proposed:
{\it e.g.} existence of the transverse flow \cite{TA,LH},
creation of small plasma droplets \cite{VH},
decays of resonances \cite{GB}, formation of the pion
system out of chemical equilibrium \cite{KR,GR},
or the medium modification of the pion dispersion
relation \cite{ES}. In this paper
we propose yet another explanation, namely that the
low-$p_T$ enhancement reflects critical scattering.

Critical scattering or critical opalescence is a rather
general phenomenon observed in the neighborhood of a second
order phase transition \cite{STANLEY}: The cross section
$d^3\sigma/d\vq$ for the scattering of light, X-rays or neutrons
on a medium is given by the expression

\begin{equalign}
\label{cr1}
{d^3\sigma  \over d\vq} \,\, \propto  \,\,
\int d^3r \int d^3r^{\prime} e^{i\vq (\vr - \vrp)}
\langle n(\vr) n(\vrp) \rangle_T,
\end{equalign}

\noindent
where $\vq$ is the momentum transfer and $\langle n(\vr) n(\vrp)
\rangle_T$ is the density-density correlation function for
the particles in the medium at the temperature $T$. In the neighborhood
of a phase transition a long-range ordering develops and
the cross section for small values of $\vq$ has the form

\begin{equalign}
\label{cr2}
{d^3\sigma \over d\vq} \,\, \propto \,\, {1 \over k^2(T) + {\vq}^2},
\end{equalign}
where
\begin{equalign}
\label{cl}
k^2(T) \,\, \propto \,\,   \mid 1 - {T \over T_c} \mid^{\gamma}
\end{equalign}

\bigskip \noindent
is the inverse correlation length with $\gamma$ being the
critical exponent.

This kind of experiment, where an {\it external} test particle is scattered
on the system in its critical state, cannot be performed for
a quark-gluon plasma which only lives for $10^{-23}$~s.
However, also particles which are created {\it inside} the plasma
should experience critical scattering and their observed momentum
distribution may bear signal of this phenomenon.
In the present paper we discuss one possible realization
of such a scenario.

In explicit calculations \cite{H1,H2}, using the Nambu -- Jona-Lasinio
(NJL) model for a description of the quark-meson plasma (no gluons),
it has been shown that singularities really occur in the cross sections
at small center-of-mass energies $\sqrt s$
of colliding particles, when the phase transition is approached.
In particular, the integrated elastic quark-antiquark cross-section
$\sigma_{q {\overline q} \rightarrow q {\overline q}}(s,T)$
diverges like $s^{-1}$ for $T \rightarrow T_c$.
A singularity is also found for the hadronization cross-section
$q {\overline q} \rightarrow \pi \pi$. In both cases the singularity
arises because the quark condensate $\langle{\overline q}q\rangle_T$,
which is the order parameter of the chiral phase transition,
goes to zero at the critical temperature.

May the observed low-$p_T$ enhancement originate from the
low $\sqrt s$ singularity in the cross sections?
We think this is so and demonstrate  it using a toy model.
Kinetic equations are proposed for the time evolution of
a spatially homogeneous quark plasma (described by the single
particle quark distribution function $\fq,\,  p=|\vp|$) into a pion gas
(with a corresponding distribution function $\fm$).
Criticality of the system enters via the singularities
in the thermalization and hadronization cross sections.

In order to keep the model as transparent as possible we work
in the relaxation time approximation, thus reducing the kinetic
equations to a coupled system of first order integro-differential
equations for $\fq$ and $\fm$. Contrary to the usual approaches,
the relaxation times $\tau (p,t)$ depend on momenta via
the energy dependence of the cross sections.
The singularities in the cross sections due to criticality of the medium
translate themselves into singularities of the inverse
relaxation times $1/\tau (p,t)$ in the momentum $p$.

\bigskip
\noindent{\large \bf Kinetic description of the hadronization of
                     the quark plasma }
\medskip

The above ideas are translated into the following set of kinetic equations

\begin{eqnarray}
\label{q}
{d\fq \over dt} &=& - {\fq - \eq \over \tth}
- {\fq \over \thd} + {\fm \over \tdc}
\\
& & \nonumber \\
& & \nonumber \\
\label{m}
{d\fm \over dt} &=& - {\fm - \em \over \tth} + {\fq \over \thd}
- {\fm \over \tdc} - {\fm \over \tem} ,
\\
& & \nonumber \\
& & \nonumber \\
\label{e}
{d\fe  \over dt} &=& {\fm \over \tem}.
\end{eqnarray}

\bigskip
Eq. (\ref{q}) determines the time evolution of the quark distribution
function $\fq$ (it describes both quarks and antiquarks, {\it i.e.}
$\fq = f_{quarks}(p,t) + f_{antiquarks}(p,t)$).
The first term on the r.h.s. of Eq. (\ref{q}) is the collision term
written in the relaxation time approximation; it is responsible for
the thermalization of quarks since the distribution function
$\fq$ is always attracted to the thermal one $\eq$.
The second (third) term describes the loss (gain) of quarks
due to the hadronization (deconfinement) process where we limit
ourselves to the reaction $q {\overline q} \rightarrow \pi \pi$
($q {\overline q} \leftarrow \pi \pi$).  Of the possible reaction
channels $q {\overline q} \rightarrow n \pi \, (n \geq 2)$, the case $n=2$
may be the dominant one for phase space reasons at moderate
energies in the $q {\overline q}$ system, and we assume that this
is so.

In Eq. (\ref{m}), for the time evolution of the distribution
functions of pions, the first three terms on the r.h.s. describe:
the thermalization of pions, appearance of pions due to
hadronization of quarks, and the two-pion reaction into
quark-antiquark pairs, respectively. The last (fourth) term
accounts for the
emission of pions from the plasma, which in a realistic system
proceeds via the surface, while we simulate it via a homogeneously
distributed sink. The emission of preequilibrium pions
is crucial in order to observe deviations from thermal equilibrium.

Eqs. (\ref{q}) - (\ref{e}) can be solved for any initial
conditions ({\it i.e.} assuming some particular form of the
distribution functions at the initial time $t=0$) provided the
thermal distribution functions $\eq$ and $\em$ are known at all times.
Since the first terms on the r.h.s. of Eq. (\ref{q}) and (\ref{m})
represent the collision terms (in the relaxation time approximation)
they must obey the symmetries leading to particle
and energy conservation. This gives the following constraints
for $\eq$ and $\em$ at each time $t$

\begin{eqnarray}    
\label{c1}
\int {\d3p \over \tp3} \,
{f_i(p,t) - f_{i,th}(p,t)\over \tth } &=& 0,
\\
& & \nonumber \\
\label{c2}
\int {\d3p \over \tp3} \, \sqrt{p^2 + m_i^2} \,\,
{f_i(p,t) - f_{i, th}(p,t) \over \tth} &=& 0
\end{eqnarray}      

\bigskip \noindent (here $i = q$ or $\pi$).
Eqs. (\ref{c1}) and (\ref{c2}) determine the temperature $T_i(t)$
and the chemical potential $\mu_i(t)$ appearing in the thermal
distributions. For simplicity we assume the low-density
high-temperature form of these functions, namely a Boltzmann
distribution

\begin{equalign}
\label{boltz}
f_{i, th}(p,t) =  g_i
\exp\left[- {\sqrt{p^2+m_i^2}-\mu_i(t) \over T_i(t) } \right],
\end{equalign}

\bigskip \noindent
where $g_i$ are the degeneracy factors:
$g_q = 24$ (quarks and antiquarks having two
different spin projections, 3 colors and 2 flavors) and
$g_{\pi} = 3$ (three different values of the isospin).


\newpage
\noindent{\large \bf Relaxation times and cross sections}
\medskip

The analysis of the exact collision term in the
Boltzmann kinetic equation (analogous to that in \cite{KH})
leads to an expression for the average time
for hadronization of two quarks into two pions

\begin{equalign}
\label{iht}
{1 \over \tau_{had}(p,t)} = {1 \over 2\sqrt{p^2+m_q^2}}
\int {d^3p_1 \over \tp3 \sqrt{p_1^2+m_q^2}}
f_q(p_1,t) F_{qq}(s) \sigma_{q {\overline q} \rightarrow \pi \pi}(s),
\end{equalign}

\noindent where $\sigma_{q {\overline q} \rightarrow \pi \pi}(s)$
denotes the total cross section
for this process. The relativistic flux factor of incoming quarks is
$F_{qq}(s) = {1 \over 2} \sqrt{s(s-4m_q^2)}$, with
$\sqrt s$ being the center-of-mass energy
of the quarks with momenta $p$ and $p_1$, respectively.
The expression giving the deconfinement relaxation time,
$\tdc$, has the form analogous to Eq. (\ref{iht}).

The relaxation time for the thermalization of quarks can be written
in the form

\begin{eqnarray}
\label{itt}
{1 \over \tau_{th}(p,t)} = & & {1 \over 2\sqrt{p^2+m_q^2}}
\int {d^3p_1 \over \tp3 \sqrt{p_1^2+m_q^2}}  f_q(p_1,t)
F_{qq}(s)
\sigma_{qq \rightarrow qq}(s)  \nonumber \\
& &  + \,\, {1 \over 2\sqrt{p^2+m_{\pi}^2}}
\int {d^3p_1 \over \tp3 \sqrt{p_1^2+m_{\pi}^2}} f_{\pi}(p_1,t)
F_{q \pi}(s)
\sigma_{q \pi \rightarrow q \pi}(s),
\end{eqnarray}

\bigskip
\noindent where the two terms on the r.h.s. of Eq. (\ref{itt})
appear due to the quark-quark and quark-pion elastic collisions
respectively. The formula for the thermalization time of pions
can be obtained formally from Eq. (\ref{itt}) by the exchange
of the indices $q$ and $\pi$.

For the sake of simplicity, in what follows we shall treat the
quarks and pions as massless particles. We shall also assume that
the cross sections:
$\sigma_{q {\overline q} \rightarrow \pi \pi}(s),
\sigma_{qq \rightarrow qq}(s),
\sigma_{\pi\pi \rightarrow \pi\pi}(s)$ and
$\sigma_{q\pi \rightarrow q\pi}(s)$
are given by the generic expression

\begin{equalign}
\label{cs}
\sigma(s) = \sigma_0 \left[1 + {s_0 \over s} \right],
\end{equalign}

\bigskip \noindent
where $\sigma_0$ represents a constant contribution to the
cross sections, and  the appearance of the singular term
$s_0/s$ accounts for the critical phenomena.
The hadronization and deconfinement cross sections,
$\sigma_{q {\overline q} \rightarrow \pi\pi}(s)$ and
$\sigma_{\pi \pi \rightarrow q {\overline q}}(s)$, are
related to each other by the principle of detailed balance

\begin{equalign}
\label{db}
g_q^2 \sigma_{q {\overline q} \rightarrow \pi \pi}(s) =
g_{\pi}^2 \sigma_{\pi\pi \rightarrow q {\overline q}}(s).
\end{equalign}

\noindent This relation guarantees that in the absence
of emission ($\tau_{em} \rightarrow \infty$), the kinetic
equations lead to the chemical equilibrium $n_q/n_{\pi}
= g_q/g_{\pi}$.

Since the relaxation
time approximation requires  that the system is always close to
thermal equilibrium, the distribution functions appearing
in Eqs. (\ref{iht} -- \ref{itt}) can be replaced by the
thermal ones and, in this way, we find the approximate
form for the hadronization time

\begin{equalign}
\label{scale}
{1 \over \thd} = {\sigma_0 n_q(t) \over 2}
\left[ 1 + {s_0 \over 4p T_q(t)} \right],
\end{equalign}
where
\bigskip \noindent
\begin{equalign}
n_q(t)=\int {\d3p \over \tp3} \fq.
\end{equalign}

\bigskip \noindent
Eq. (\ref{scale}) can be used to find the characteristic
momentum scale, $\delta p(t) = s_0/4T_q(t)$, below which the
hadronization time significantly deviates from a constant.
Due to the faster hadronization of low-energy quarks their
temperature $T_q(t)$ increases and, correspondingly, $\delta p(t)$
decreases in time. Therefore, in the following we shall use
the time average of $\delta p(t)$, {\it i.e.} the quantity
$\delta p = \langle \delta p(t) \rangle$, in order to make
the estimate of the momentum range, $0 < p < \delta p$,
where the enhancement in pion production is expected.

In our model the pions are emitted
from the whole volume of the interacting system.
The most natural physical assumption for $\tem$ would be,
that it is inversely proportional to the velocity of the pion,
which for massless particles equals $c$ for all momenta.
For this reason we choose $\tau_{em}$ to be a constant.
In order to study the sensitivity of our results on the
emission rate we shall introduce the ratio $\cal R$ defined as

\begin{equalign}
\label{R}
{1 \over \tau_{em}}= \, {\cal R} \,\,
{1 \over \tau_{th}(p \rightarrow \infty, t = 0)}.
\end{equalign}

\bigskip
\noindent
The magnitude of $\cal R$ indicates how much faster
the evaporation is, in comparison to the rate of the
thermalization processes inside the system.

\newpage
\noindent{\large \bf Results}
\bigskip

In order to reduce the number of free parameters in
our calculation we have set all the cross sections:
$\sigma_{q {\overline q} \rightarrow \pi\pi}(s),
 \sigma_{qq \rightarrow qq}(s),
 \sigma_{q \pi \rightarrow q \pi}(s)$ and
$\sigma_{\pi\pi \rightarrow \pi\pi}(s)$ to be equal
(within a factor of 3 such a result is supported by
calculations within the NJL model). Then, we are left with
three parameters: $\sigma_0, s_0$ and $\cal R$.
One can notice that $\sigma_0$ sets the overall
time-scale and is unimportant if we are interested
in the final $(t \rightarrow \infty$) pion distributions.

We start solving our kinetic equations by assuming that
the initial quark distribution is a thermal one and that
there are no pions in the system. After integrating
Eqs. (\ref{q}) -- (\ref{e}) till the time when all
quarks are hadronized and all pions are emitted, one
obtains the distribution function $f_{em}(p,t \rightarrow
\infty)$ of the observed pions for a set of parameters
$s_0$ and $\cal R$.

In Fig. 1 we show our results for the case when the initial
temperature of the quarks $T_q(t=0)$ = 140 MeV. The dashed
lines represent the initial quark distribution functions,
whereas the solid ones represent the distribution
functions of the emitted (observed) pions for four
different choices of the parameters:
(a) ${\cal R} = 33, s_0 = 0.48$ GeV${}^2$;
(b) ${\cal R} = 11, s_0 = 0.48$ GeV${}^2$;
(c) ${\cal R} = 33, s_0 = 0.16$ GeV${}^2$; and
(d) ${\cal R} = 11, s_0 = 0.16$ GeV${}^2$.
The corresponding values of the momentum scale
$\delta p$ are: 210, 260, 90 and
95 MeV, respectively.

In all the considered cases one can clearly see
the enhancement in the pion production for
$ p < \delta p$. During the evolution of our
system the low energy quarks hadronize more
effectively because of the $p$-dependence of
$\tau_{had}$ and, correspondingly, the low energy pions
are preferably produced. Our results also show
that the effect depends strongly on the pion
emission rate: for larger values of $\cal R$ the
effect is more significant. Such a dependence
can be easily understood since the thermalization
leads always to the distribution functions of the
form (\ref{boltz}), and in the limiting case
${\cal R} \rightarrow 0$ no effect could be seen.
The last fact indicates also that our mechanism
for obtaining the low-$p_T$ enhancement is
basically a non-equilibrium one: The excess
can be observed only if the pions are emitted from a
non-equilibrium system.

In Fig. 2 we plot the ratio $r(p)$ obtained by
normalization of the distribution function of the
observed pions to the Boltzmann distribution.
In this case the latter is obtained by fitting
the exponent function to $f_{em}(p,t \rightarrow \infty)$
in the region 0.5 GeV $< p <$ 1 GeV.
The values of the parameters are the same as those
assumed in Fig. 1a. Our result is shown together
with the data \cite{STACHEL} (14.6 A GeV
Si + Pb $\rightarrow \pi^{-}$, rapidity interval
y = 3.4 -- 3.6). Because we did the calculations in
the special case $m_{\pi} = 0$, we treat the data as
if they were obtained also for massless pions and identify
the quantity $m_T - m_{\pi}$ (appearing in \cite{STACHEL})
with the transverse momentum $p_T$. Moreover, using
the relation $p_T = \sqrt{2p/3}$ (valid for isotropic systems)
we find $m_T - m_{\pi} = \sqrt{2p/3}$.

Fig. 2 shows that our model is
able to explain a characteristic shape of the observed
enhancement, if the two parameters ($s_0$ and ${\cal R}$)
are adjusted. Nevertheless, we are of the opinion that
the low-$p_T$ enhancement in pion production is caused by the
superposition of various mechanisms.
In Ref. \cite{STACHEL} it is argued that the
decay of $\Delta$ resonances can be responsible for
the excess of low-$p_T$ pions, although in different
rapidity intervals different ratios of pions from
$\Delta$ decay to direct pions must be assumed
to fit well the data. This uncertainity leaves room
for other effects. Also the origin of the
enhancement in kaon production, starting at much
smaller values of $p_T$, is still not clear.
Therefore, we think that our mechanism
for obtaining the low-$p_T$ enhancement is probably
only one of a few contributions leading to this effect.

\bigskip
\noindent{\it Acknowledgments:}  We thank Sandy Klevansky
for critical comments concerning the manuscript.

\noindent {\bf Figure Caption}

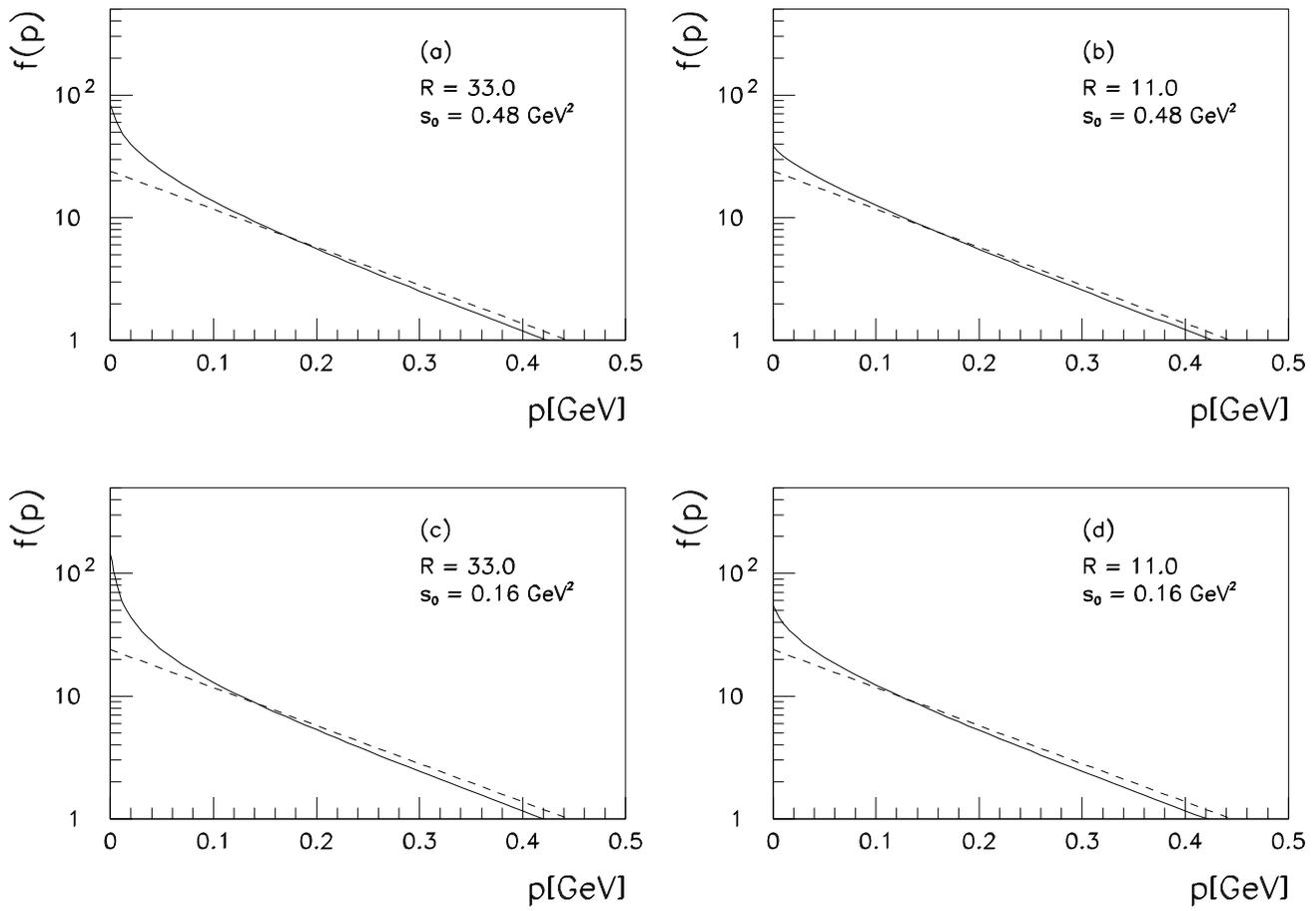
{\bf Fig. 1} The initial quark distribution functions
(dashed lines) and the distribution functions of the
emitted pions (solid lines) for four different
choices of the parameters:
(a) ${\cal R} = 33, s_0 = 0.48$ GeV${}^2$;
(b) ${\cal R} = 11, s_0 = 0.48$ GeV${}^2$;
(c) ${\cal R} = 33, s_0 = 0.16$ GeV${}^2$; and
(d) ${\cal R} = 11, s_0 = 0.16$ GeV${}^2$.

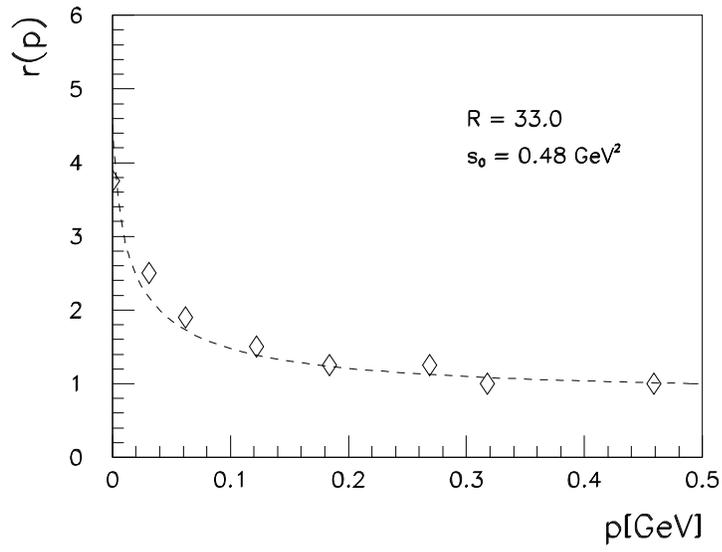
{\bf Fig. 2} The distribution function of the observed
pions normalized to the Boltzmann distribution.
The parameters as in Fig. 1a. The diamonds show the
data \cite{STACHEL}.


\newpage

\begin{thebibliography}{99}


\bibitem{SS} H.R. Schmidt and J. Schukraft, J. Phys. {\bf G19}
(1993) 1705.

\bibitem{QM} Proceedings of the conference {\it Quark Matter '93},
Nucl. Phys. {\bf A566} (1994).

\bibitem{TA} T.W. Atwater, P.S. Freier and
J.I. Kapusta, Phys. Lett. {\bf B199} (1987) 30.

\bibitem{LH} K. Lee, U. Heinz and
E. Schnedermann, Z. Phys. {\bf C48} (1990) 525.


\bibitem{VH} L. Van Hove, CERN preprint, TH-5236/88 (1988).

\bibitem{GB} G.E. Brown, J. Stachel and
G.M. Welke, Phys. Lett. {\bf B253} (1991) 19.

\bibitem{KR} M. Kataja and P.V. Ruuskanen, Phys. Lett. {\bf B243}
(1990) 181.

\bibitem{GR} S. Gavin and P.V. Ruuskanen, Phys. Lett. {\bf B262}
(1991) 326.

\bibitem{ES} E. Shuryak, Phys. Rev. {\bf D42} (1990) 1764.

\bibitem{STANLEY} H.E. Stanley, {\it Introduction to Phase
Transitions and Critical Phenomena} (Oxford University Press,
New York, 1971).

\bibitem{H1} P. Zhuang, J. H\"ufner, S.P. Klevansky, and
L. Neise, Heidelberg preprint HD-TVP-94-09.

\bibitem{H2} J. H\"ufner, S.P. Klevansky, E. Quack and
P. Zhuang, Heidelberg preprint HD-TVP-94-13, to appear
in Phys. Lett. {\bf B}.

\bibitem{KH} K. Huang, {\it Statistical Physics}
(Wiley, New York, 1987), Sect. 5.4.

\bibitem{STACHEL} J. Stachel, Nucl. Phys. {\bf A566} (1994)
183c.

\end{thebibliography}
\end{document}